\documentclass[article,twocolumn,showpacs,superscriptaddress,groupedaddress]{revtex4-1}  

\usepackage{graphicx}  
\usepackage{dcolumn}   
\usepackage{bm}        
\usepackage{amssymb}   
\usepackage{amsmath}
\usepackage{flushend}
\usepackage{color}
\usepackage{scrextend}

\newcommand{\beq}{\begin{equation}}
\newcommand{\eeq}{\end{equation}}
\newcommand{\bea}{\begin{eqnarray}}
\newcommand{\eea}{\end{eqnarray}}

\newcommand{\nn}{\nonumber\\}

\newcommand\eqr[1]     {(\ref{#1})}

\newcommand\fig[1]     {Fig.\,{\ref{#1}}}

\begin{document}

\title{Critical scaling in the large-$N$ $O(N)$ model in higher dimensions and its possible connection to quantum gravity}
\author{P. Mati}

\email{Peter.Mati@eli-alps.hu}
\email{matipeti@gmail.com}
\affiliation{ELI-ALPS, ELI-Hu NKft, Dugonics t\'er 13, Szeged 6720, Hungary}
\affiliation{MTA-DE Particle Physics Research Group, P.O.Box 51, H-4001 Debrecen, Hungary}

\date{\today}

\begin{abstract} 
The critical scaling of the large-$N$ $O(N)$ model in higher dimensions using the exact renormalization group equations has been studied, motivated by the recently found non-trivial fixed point in $4<d<6$ dimensions with metastable critical potential. Particular attention is paid to the case of $d=5$ where the scaling exponent $\nu$ has the value $1/3$, which coincides with the scaling exponent of quantum gravity in one fewer dimensions. Convincing results show that this relation could be generalized to arbitrary number of dimensions above five. Some aspects of AdS/CFT correspondence are also discussed.\\ \\
{\textit {In loving memory of my grandmother.}}
\end{abstract}

\maketitle

\section{Introduction}
The non-trivial critical behavior in $O(N)$ theories are well-known for dimensions $d<4$ \cite{on}. 
Thus, a statement on the existence of interacting critical theories beyond four space-time dimensions is rather 
unusual since one would expect the triviality of the $O(N)$ vector model in general \cite{triv}.
However, in recent works \cite{kleb1,kleb2,Gracey}, exhaustive one, three and four loop analyses of the $O(N)$ theory with cubic interactions and $N+1$ scalars show that the large-$N$ $O(N)$ theory could follow the asymptotically safe 
scenario under the renormalization group in the UV. More precisely, it was argued that the IR fixed point found in the aforementioned $O(N)$ theory with the cubic interaction is equivalent to a perturbatively unitary UV fixed point 
in the large-$N$ $O(N)$ model for dimensions  $4 < d < 6$. The presence of such UV fixed point could be particularly interesting due to the conjectured AdS$_{d+1}$/CFT$_d$ duality between a higher-spin $d+1$-dimensional massless gauge theory in AdS space (with an appropriate boundary condition) and the large-$N$ critical $O(N)$ model in $d$ dimensions \cite{hs}. The former is called the Vasiliev theory, which describes a minimal interacting theory with gravity and higher-spin fields in its spectrum. It can be obtained as the tensionless limit of string theory, where the infinite tower of higher-spin string modes are massless, and since there is no energy scale it can be considered as a toy model describing physics beyond the Planck scale \cite{vas}. Studies related to the existence of the UV fixed point in the large-$N$ $O(N)$ model, using conformal bootstrap approach and exact (or functional) renormalization group (ERG or FRG) methods, can be found in \cite{Rey,boot1,boot2} and \cite{perc1,mati1}, respectively.\\
The structure of the paper is the following. In section \ref{uvfp} a discussion of the previous results on the UV fixed point are given. It is shown that the solution by polynomial expansion coincides with one of the infinite many solutions (the physically most sensible one) of the exact treatment. In section \ref{criti} the non-trivial critical scaling for higher dimensional theories is derived. In \ref{gravv} a speculation on the possible connection to quantum gravity is presented. 

\section{Identifying the UV fixed point}\label{uvfp}
First, a brief review of the analytical
results from \cite{perc1} is given. Let us consider the effective average action of the O($N$) symmetric theory in $d$ dimensions within the local potential approximation (LPA):
\beq\label{Gamma}
\Gamma_k=\int d^d x \left[ \frac{1}{2}(\partial\bar\phi)^2 + U_{k} (\bar\phi^2) \right].
\eeq
$U_k$ is the dimensionful potential depending on $\bar
\phi^2$, where $\bar\phi$ is the dimensionful vacuum expectation value (VEV) of the field. The subscript $k$ stands for the 
RG scale i.e., the Wilsonian cutoff,  which defines the effective theory. In the large-$N$ limit the anomalous dimension of the Goldstone modes disappears, therefore, setting the wave function renormalization constant to unity in \eqref{Gamma} gives a well-justified approximation. In fact, the LPA is considered to be exact in the large-$N$ limit of the $O(N)$ model \cite{zinn,zinn2,analy}.
The flow of the effective action is given by the exact functional differential equation
\cite{FRGgen}
\begin{equation}\label{flow}
\partial_t\Gamma_k=\frac{1}{2}\text{Tr}\left(\Gamma _k ^{(2)} + R_k \right)^{-1} \partial_t R_k.
\end{equation}
Here, the logarithmic flow parameter $t=\ln ({k}/{\Lambda})$ (where $\Lambda$ is the initial UV scale) 
is introduced with a momentum dependent regulating function $R_k(q^2)$ which ensures that the fluctuations above the Wilsonian cutoff scale are integrated out.
$\Gamma _k^{(2)} [\bar\phi]$ is used as a shorthand notation for the second derivative with respect to the field and the trace denotes the 
integration over all momenta as well as the summation over internal indices.
This integral is evaluated by choosing $R_k(q^2)$ so that $\Gamma_k$ approaches the bare action in the limit $k\rightarrow\Lambda$ and the full quantum effective action when $k\rightarrow 0$ \cite{FRGgen}. A detailed study of an extensive class of regulator functions is reported in  \cite{CSS}. In the current case, the optimized regulator is chosen $R_{k}(q^2)=(k^2-q^2)\,\theta(k^2-q^2)$ which provides an analytic result for the momentum integral \cite{opt_rg}. It is convenient to introduce $\bar\rho\equiv\frac{1}{2}\bar\phi^2$, which will be used throughout this paper. Inserting \eqref{Gamma} into \eqref{flow} and applying the limit $N\to\infty$ yields the flow for the effective potential in the large-$N$ \cite{analy}:
\begin{equation}\label{flow2}
\partial_t u= - d u +(d-2)\rho u'+ \frac{1}{1+u'},
\end{equation}
where the dimensionless quantities $u=Uk^{-d}$ with $\rho=\bar\rho k^{-d+2}$ are introduced and $u'=\partial_\rho u$.
\par
\begin{figure}[h!]
	\includegraphics[width=.42\textwidth]{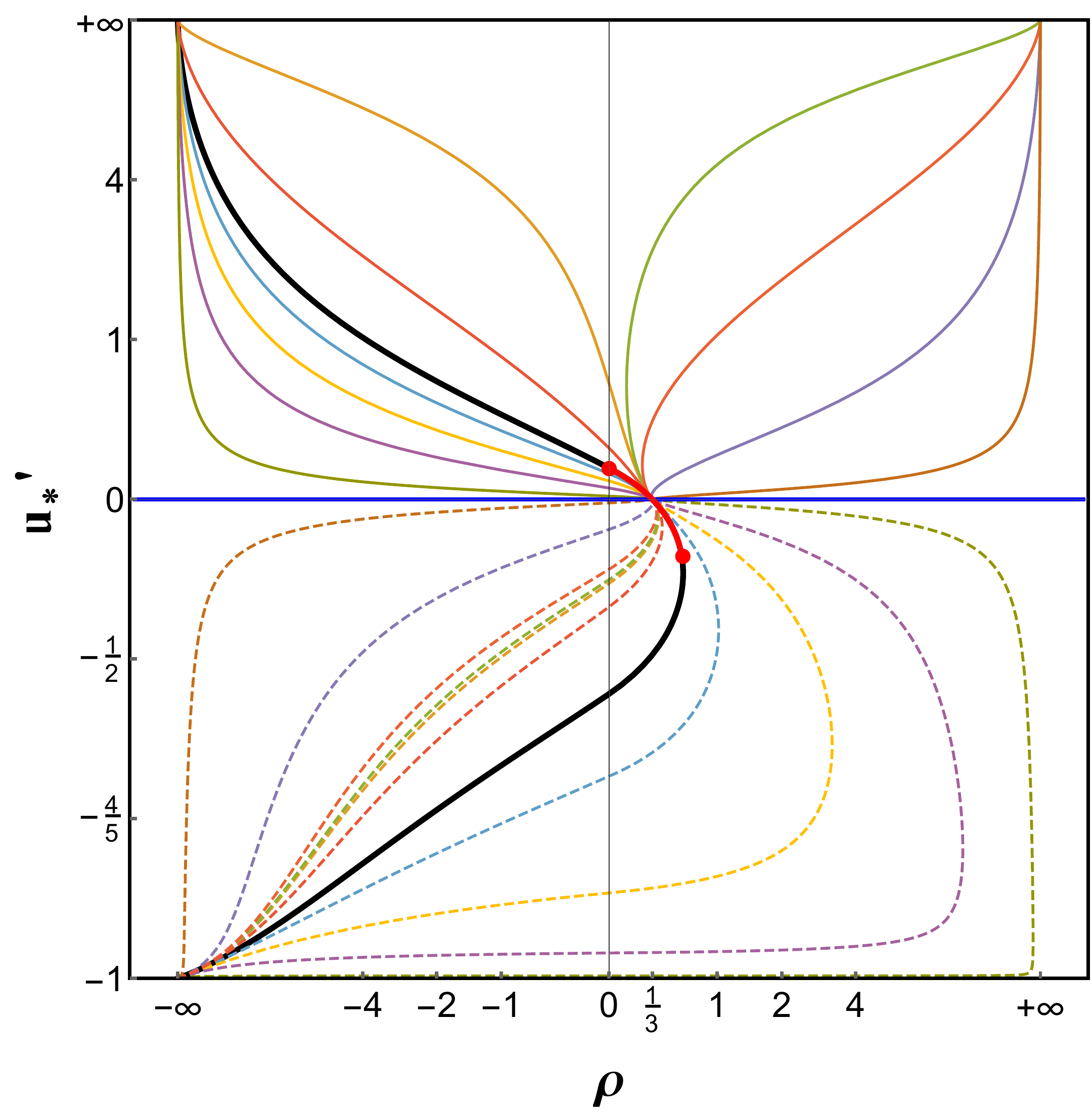}
	\caption{Each curve corresponds to a critical potential derivative with a particular choice of 
		$c$ in $d=5$. The thick black line is for the $c=0$ solution. The red line segment is considered as the physical branch. The axes are rescaled for clarity.}\label{uprim}
\end{figure}

 An exact solution for the $\rho$ derivative of \eqref{flow2}  can be obtained by using the method of characteristics \cite{analy, Marchais} and the fixed point solutions associated to \eqref{flow2} can be given as an implicit function $\rho=\rho(u_*')$. $u_*$ is introduced as the dimensionless effective potential at the fixed point. The most compact form of these exact solutions  for $d=2n+1$ and $d=2n$ ($n\in {\mathbb{Z}}$) are respectively
\beq\label{oddd}
 \rho=c u_*'^{\frac{d}{2}-1}+\frac{1}{(d-2)}\, _2F_1\left(2,1-\frac{d}{2};2-\frac{d}{2};-u_*'\right) 
\eeq
and
\beq\label{even}
\rho=\bar c u_*'^{\frac{d}{2}-1}+\frac{1}{(d+2)(1+u_*')^2}\, _2F_1\left(1,2;2+\frac{d}{2};\frac{1}{1+u_*'}\right),
\eeq
where $c$ is an arbitrary constant obtained from the integration, $\bar c=c-\frac{d\pi}{4}\sin(d\pi/2)$ and $_2F_1$ is the hypergeometric function. \fig{uprim} shows the solutions for the case in $d=5$. Each curve corresponds to a solution with a particular value of the parameter $c$.
$u_*' \geq0$ \eqref{oddd} holds for every $c\in \mathbb{R}$ but in order to obtain a continuation of the solutions to $u_*'\leq0$ the constant $c$ needs to take imaginary values.
There is one exception: the solution corresponding to $c=0$.
This is shown in \fig{uprim} as the thick black curve that passes smoothly through $u_*=0$ and intersects the horizontal and vertical axes at $\rho=1/3$ and $u_*'(0)\approx0.1392$ on the upper plane, respectively.
It is tempting to consider this fixed point potential as the physical one since it is analytic at its extremum. On the other hand, this curve still has the problem like the other solutions (including their continuation): $u_*'$ can be only considered as a multivalued function of $\rho\in[0,0.6214]$ \cite{perc1}.
\par Focus now shifts to the author's previous results \cite{mati1} where a different technique, based upon polynomial expansion, was used to find the fixed point solutions of the flow equation. The potential is assumed to be analytic in this case:
\begin{eqnarray}
u(\rho)=\lim\limits_{n\to\infty}\sum\limits_{i=1}^{n} \frac{u^{(i)}(0)}{i!}\rho^i.
\end{eqnarray}
The derivatives of the potential are the couplings of the theory: $u'(0)=g_1=m^2$ 
(squared mass), $u''(0)=g_2=\lambda$ (quartic coupling), etc... An efficient algorithm was worked out for finding the fixed points of the theory for expansions up to order 50, if required. This method is based on the observation that all the couplings can be expressed through the squared mass of the system at the fixed points, $g_i^*=g_i^*(m_*^2)$.

\begin{figure}[h!]
	\includegraphics[width=.47\textwidth]{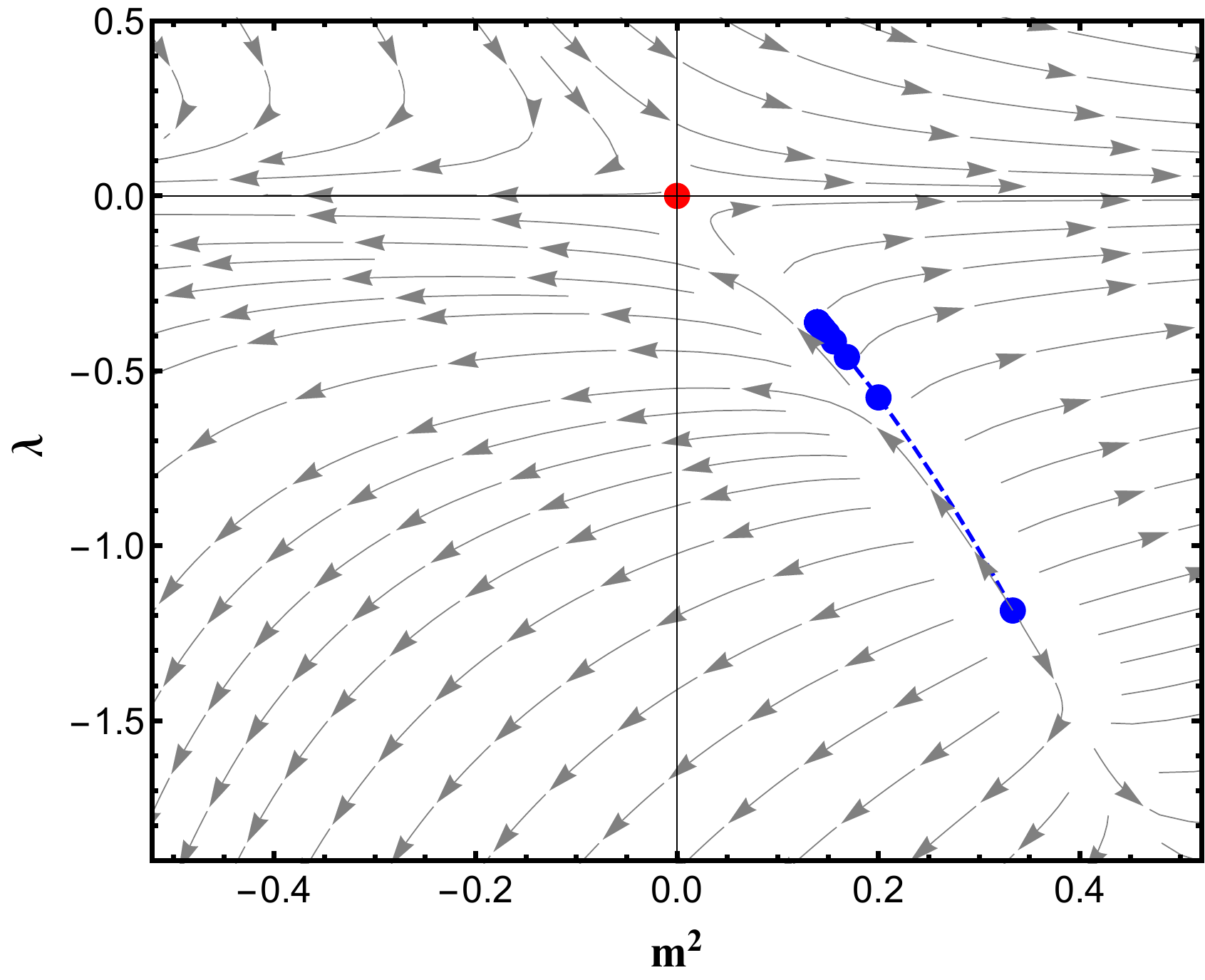}
	\caption{The RG flow in the $\{m^2,\lambda\}$ hyperplane of theory space. The drifting of the non-trivial UV fixed point is shown towards $\{m_*^2,\lambda_*\}\approx\{0.1392,-0.3613\}$ with increasing expansion order (blue dots). The red dot at the origin represents the GFP. Arrows point from UV to IR scales.}
	\label{flow fig}
\end{figure}   

\noindent In the case of the five-dimensional large-$N$ $O(N)$  model, the fixed point structure shows a non-trivial fixed point as well as the non-interacting Gaussian fixed point (GFP).
The fixed point position drifts as the order of the Taylor-expansion increases and converges to $m_*^2\approx 0.1392$ as shown in \fig{flow fig}.
As $m^2=u'(0)$, it can be confidently stated that this is the same fixed point solution as found by the analytic study of the flow when $c=0$ in \eqref{oddd}.
In fact, this technique naturally singles
out a fixed point solution from all the other solutions that are present only in the analytical case. In \fig{uprim} this corresponds to the red line segment on the thick black curve and will be considered as the physical solution.
\par
\begin{figure}[h!]
	\includegraphics[width=.45\textwidth]{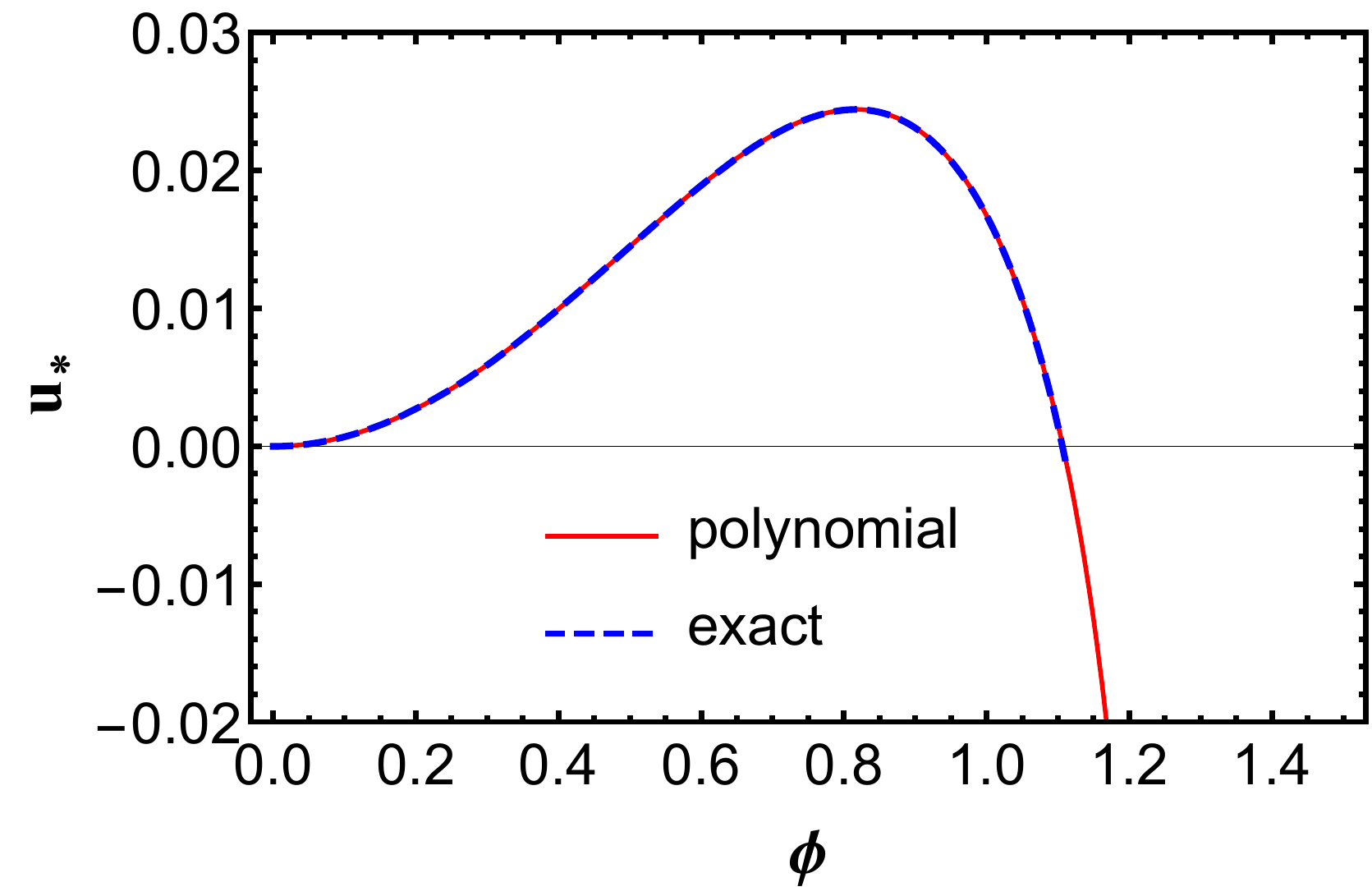}
	\caption{The metastable critical potential in $d=5$. On the horizontal axis the field VEV ($\phi=\sqrt{2 \rho}$) was used hence the exact potential is only valid between $\phi\in[0,1.1148]$.}\label{pot}
\end{figure}

 \fig{pot} shows the comparison between the exact critical potential, computed from \eqref{oddd}, and the polynomial expansion results, $u_*(\rho)=\sum_{i=1}^{n}\frac{g^*_i(m_*^2)}{i!}\rho^i$, with $n=26$. The non-analytic nature of the exact potential is very apparent as it is restricted to the finite interval unlike to the polynomial potential which was assumed to be analytical. The most important features of this potential are the metastable ground state and the lack of a true 
vacuum. This may initially discourage further investigations, but metastable and unstable vacua are not unknown. There is the question on the electroweak vacuum stability for instance: it is still a topical question if the Higgs potential exhibits a ground state or if there is an unstable universe existing in a false vacuum (with a very long lifetime) \cite{Higgsgen}. Alternatively, the theory with the metastable potential could be saved from the AdS side, too. As mentioned above, the critical large-$N$ $O(N)$ theory in $d=5$ is possibly dual to a Vasiliev higher-spin theory in AdS$_6$ space, thus they must have the same energy spectrum. In AdS space, the Breitenlohner-Freedman (BF) bound gives a negative, dimension dependent lower bound for the squared mass of the field, above which the theory can be considered as stable \cite{BFree}. 

\noindent The BF bound can also be generalized for massless higher-spin fields that also depends on the spin value \cite{BFreemless}. 
In turn, the same argument could hold for the other branch of $u_*'(\rho)<0$ for $\rho\in[0,0.6214]$ with $u_*'(0)=m_*^2\approx-0.5776$, \fig{uprim}. In this case the potential is completely unstable in the restricted interval as $u'(\rho)<0$. However, this fixed point potential is completely ignored by the polynomial approach.

\section{Critical scaling}\label{criti}
Despite the unconventional properties of the potential, it is still possible to extract the critical exponent $\nu$.
This is the scaling exponent of the correlation length (or inverse mass) and characterizes the system at criticality.
For the exact determination of the exponent in $d$ dimensions the method of eigenperturbation is used, which is based on the linearized flow around the fixed point, i.e. 
$u(\rho,t)=u_*(\rho)+\delta u(\rho,t)$ \cite{Marchais,eigp}. Using \eqr{flow2} the fluctuation equation for the derivative of the potential reads:
\beq
\partial_t\delta u' = 2 \frac{u_*'}{u_*''}\left(\partial_\rho - \frac{(u_*' u_*'')'}{u_*'u_*''}-\frac{d-4}{2}\frac{u_*''}{u_*'} \right)\delta u'.
\eeq
This can be thought of as an eigenvalue problem: $\partial_t\delta u' =\theta \delta u'$, where the smallest eigenvalue $\theta$ equals the negative inverse of the scaling exponent $\nu$. 
Solving this PDE via the method of separation of variables yields
\bea\label{pert}
\delta u' \propto e^{t \theta}u_*'^{\frac{1}{2}(\theta + d-2)}u_*''.
\eea
The details of this computation are provided in the Appendix.
Perturbations at the node ($u_*'(\rho_0)=0$) are required to have a high regularity so restrictions on the values of $\theta$ are necessary in order to keep $\delta u'$ analytic.
Both formulae in \eqr{oddd} and \eqr{even} at $u_*'=0$ take the value $\rho(0)\equiv\rho_0=1/(d-2)$. Using Taylor expansion around $\rho_0$, and setting $c=0$ and $\bar c=0$, a linear behavior of $u_*'$ can be found, $u_*'\propto \left(\rho-\frac{1}{d-2}\right)$.
This makes $u_*''$ a constant and substituting back this expression into \eqr{pert} gives
\beq
\delta u' \propto e^{t \theta}\left(\rho-\frac{1}{d-2}\right)
^{\frac{1}{2}(\theta + d-2)}.
\eeq
The allowed values are then $\theta= 2 (l+1-d/2)$, where $l$ is a non-negative integer, and the scaling exponent
is obtained by the lowest value of $\theta$, i.e. for $l=0$. Thus, the scaling exponent for arbitrary dimensions in the large-$N$ $O(N)$ model is 
\beq\label{nuu}
\nu=(d-2)^{-1}.
\eeq
By using the polynomial expansion, the critical exponent $\nu$ can be calculated as the negative inverse of the lowest eigenvalue of the stability matrix at the fixed point $B_{ij}=\left.\partial \beta_i /\partial g_j\right|_{g=g_*}$ \cite{FRGgen}, where the beta functions are defined as the RG scale derivative of the couplings: $\beta_i=\partial g_i(t)/\partial t$.
As the LPA became exact in the large-$N$ limit, the correct value for the critical exponent can be obtained at every order of the expansion, i.e. \eqr{nuu} for arbitrary dimensions.
This relation, on the other hand, is well-known for the large-$N$ $O(N)$ theories in $d\leq4$ \cite{zinn, zinn2, kardar, Cardy}. However, it was not extended to higher dimensions as the upper critical dimension was considered to be $d=4$.
Yet, with an accurate analysis of the fixed point structure for $d>4$, it seems that a non-trivial fixed point can be found in the UV, where, instead of the mean-field scaling, the relation \eqr{nuu} still holds. However, the effective potentials defined at criticality are non-analytic and/or metastable for these values of $d$ and care must be taken with interpreting these results.
In particular, in five dimensions $\nu=1/3$ and the ground state seems to be metastable.
In the papers \cite{kleb1,kleb2} also an unbounded critical potential is expected, and in that respect, the results presented here are consistent with those.
\par
\begin{figure}[h!]
	\includegraphics[width=.425\textwidth]{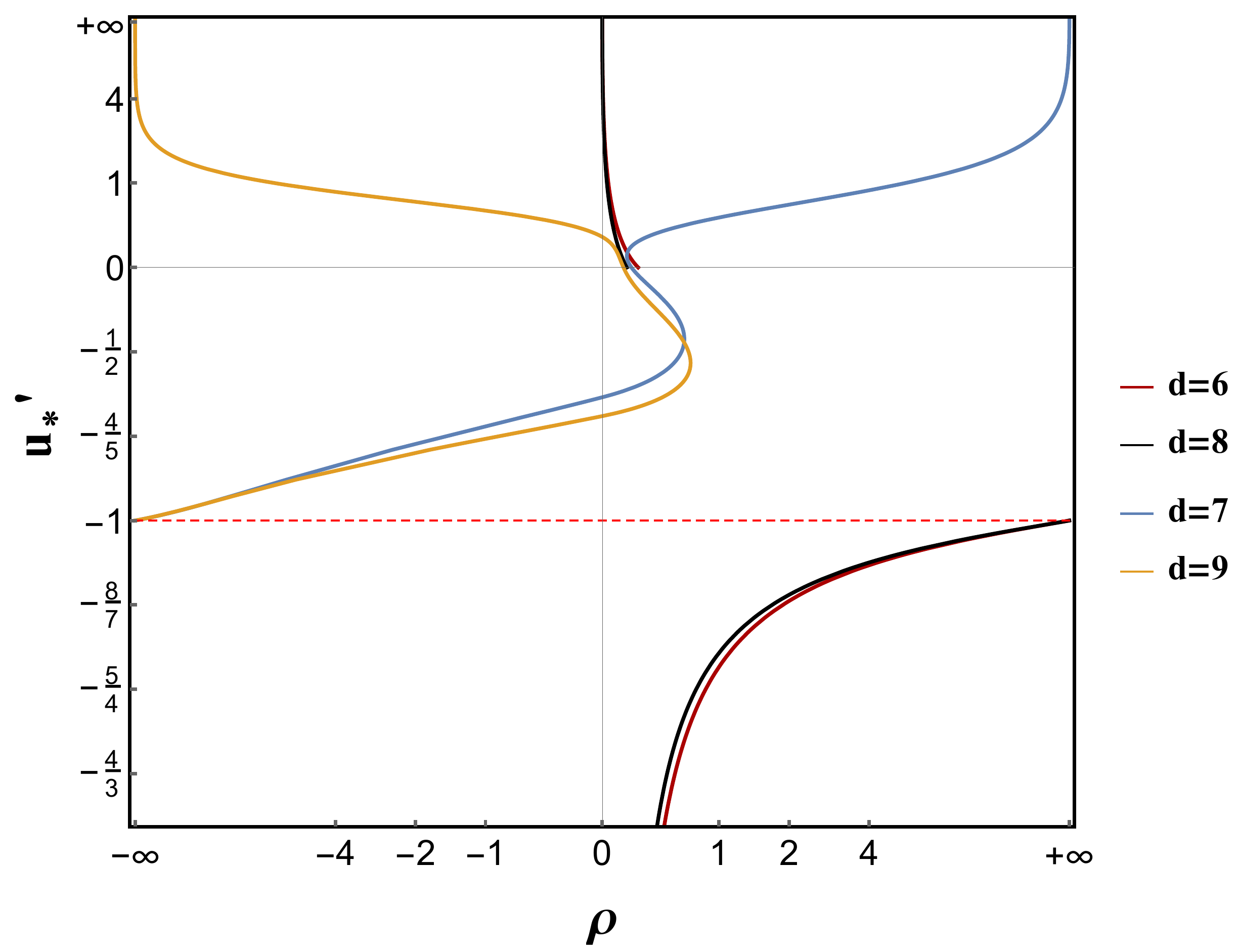}
	\caption{The fixed point solutions given by \eqref{oddd} and \eqref{even} in $d\geq6$ for $c=0$ and $\bar c=0$, respectively. The axes are rescaled for clarity.}
	\label{higher}
\end{figure}

 Although in the large-$N$ limit the dimensionality is restricted to $4<d<6$ due to the unitarity bound \cite{kleb1,kleb2,unitarity1}, higher dimensional cases can be also studied.
\fig{higher} displays the solutions \eqref{oddd} and \eqref{even} for $d>5$ with $c=0$ and $\bar c =0$, respectively. The following observations can be made. In $d=6, 8$ dimensions $u_*'$ is singular at $\rho=0$ and multivalued for $\rho<\rho_0$, in addition, the function \eqref{even} also gets complex for $u_*'\in[0,-1]$ making the theory non-unitary. In $d=7$ the potential seems to be stable but, because of the turning points, it becomes multivalued, although at $\rho=0$ it is unique.
In $d=9$ the situation is very similar to \fig{uprim} for the $c=0$ case.
These three categories seem to be preserved to all even, $d=4n+3$ and $d=4n+1$ ($n\geq1$) dimensions, respectively.
In the even-dimensional case, it is very hard to give a physical interpretation due to its singular structure and complex nature. For the $d=4n+3$ cases, three branches can be defined due to the "S" shape of the curve around $u'_*=0$, making it challenging to understand its physical content.
Perhaps certain parts of the "S" shape could be removed in the spirit of Maxwell's construction \cite{max}, which would allow us to define a bounded but non-analytic function.
When $d=4n+1$, the same arguments as in $d=5$ case can be used to define a metastable potential.

\noindent It is also worth mentioning that a similar convergence to \fig{flow fig} can be observed clearly only for $4<d<6$ ($d\in\mathbb{R}$) by using the polynomial approximation.
From the analytical side, using \eqref{oddd}, these solutions have the same structure as in \fig{uprim}.
This might suggest that physically sensible fixed points exist in $4<d<6$, provided that metastability is accepted.
However, although the relation for the scaling exponent $\nu$ holds naively for all $d$, further investigations are required for the $d\geq6$ cases both for integer and fractal dimensions. A recent study related to higher dimensional $O(N)$ theories can be found in \cite{Gracey2}.

\section{ On the possible connection to quantum gravity}\label{gravv}

Attention now switches to an interesting observation which might link the large-$N$ $O(N)$ model to quantum Einstein gravity (QEG). Much of the current evidence suggests that QEG admits a continuous phase transition between physically two distinct phases described by a strong and weak Newton's coupling \cite{QEG1}.
This phenomenon is naturally associated to a UV fixed point which
is characterized by a non-trivial scaling of the correlation length: $\xi\propto \left|G_b-G_* \right|^{-\nu}$, where the dimensionless quantities $G_b$ and $G_*$ are the bare and the fixed point Newton's coupling, respectively.
Within the framework of FRG in \cite{Falls}, using the optimized regulator and a special reparametrization of the metric fluctuation that ensures the gauge independence, $\nu^{-1}=-6 + 4/d + 2 d$ can be obtained.
Substitution of $d=4$ results $\nu=1/3$.
The scaling exponent $\nu\simeq1/3$ has been found by using the Regge lattice action in Hamber's extensive numerical studies \cite{Hamber3,Hamber4,Hamber5}.
A simple geometrical argument is given in support of the exact value of $1/3$ \cite{Hamber4}.
It is based on the observation that the quantum correction to the static gravitational potential, due to the vacuum-polarization induced scale dependence of Newton's coupling, can be interpreted as a uniform mass distribution surrounding the original source only if $\nu^{-1}=d-1$ for $d\geq4$.
In particular, for $d=4$ this gives  $\nu=1/3$.
This conjecture can be compared to the results obtained in \cite{Falls} by inserting different values for $d$ greater than four: $\nu(d=5)\approx0.2083$, $\nu(d=6)=0.15$.
Moreover, these $\nu$ values might improve by taking into account higher order curvature invariants in the effective action.
These results suggest that an interesting relationship can be revealed between the critical exponents of the large-$N$ $O(N)$ model ($\nu_O$) and QEG ($\nu_G$) as a function of dimension:
\beq\label{onqeg}
\nu_O(d)\simeq\nu_G(d-1),\qquad\text{for }d\geq5. 
\eeq
A similar phenomenon in critical systems, called Parisi-Sourlas dimensional reduction, shows that particular classical field theories and a corresponding quantum field theory in two fewer dimensions could fall into the same universality class \cite{dimred,dimred5,dimred6,branched1,branched2}. It is highly non-trivial whether the underlying mechanism is the same in the present case, however, two dimensional difference can be found between the classical Vasiliev theory and quantum gravity. Another observation could further support the interesting relation which is conjectured by \eqref{onqeg}: both QEG and the large-$N$ $O(N)$ model can be related to branched polymer systems. Being more precise, it is widely believed that QEG is described by a branched polymer-like system in its weakly coupled phase \cite{Hamber3,Hamber4,Hamber5,GravPolymer}. Similarly, $O(N)$ models represent discretized branched polymers at the double scaling limit (i.e. when $N\to\infty$ and $g\to g_*^i$ in a correlated manner) \cite{zinn2,OnPolymer}. In particular, for $d=5$ equation \eqref{onqeg} can be considered to be exact (provided that the QEG exponent is exactly $1/3$), and as it is pointed out in \cite{Hamber3}, the critical exponent $\nu=1/3$ possibly corresponds to a branched polymer system: in $d=4$ the exponent $\nu_p=1/2$ and at the upper critical dimension $d=8$, $\nu_p=1/4$ (where the lower index $p$ stands for 'polymer'). One would expect a branched polymer system with $\nu_p=1/3$ for $d\in(4,8)$. Another interesting remark can be made by considering the results of \cite{Gracey2} where also some interdimensional universality is shown between different field theories. Considering all these results, it might be possible that a more fundamental connection is emerging in the $d$-dimensional view of these various theories. 
Despite the relation found in \eqref{onqeg} the two theory does not necessarily fall into the same universality class, unless there is way to relate all the critical exponents.
There is already a conflict between the most conventional value of the anomalous dimension of the graviton in QEG ($\eta_G=-2$) and $\eta_O=0$.
However, if the usual scaling laws \cite{kardar} are assumed to be valid in QEG, $\eta_G=-2$ in $d=4$ gives $\delta_G\to\infty$, which is rather questionable for a critical exponent.
It would be of interest to find out if the relationship described in \eqref{onqeg} is a mere coincidence or if there is a deeper explanation that implies a correspondence between QEG$_{d-1}$ and the large-$N$ $O(N)$ theory in $d$ dimensions which is in turn dual to the higher-spin Vasiliev theory in AdS$_{d+1}$ space (where $d\geq5$).

\section*{Acknowledgement}
The author would like to thank G. S\'arosi, A. Jakov\'ac and Zs. Sz\'ep for the very useful discussions and their comments on the manuscript. The author also would like to thank K. Falls the discussions on quantum gravity. The ELI-ALPS project (GOP-1.1.1-12/B-2012-0001) is supported by the European Union and co-financed by the European Regional Development Fund. This research has been also supported by the Hungarian Science Fund under the contract OTKA-K104292.

\appendix
\setcounter{secnumdepth}{0}
\section{Appendix}
In the following the derivation of the eigenperturbation is presented in details. Differentiating \eqref{flow2} with respect to $\rho$ yields:
\beq\label{derivfl}
\partial_t u' = - 2u' + (d-2)\rho u''-\frac{u''}{(1+u')^2}.
\eeq
The $u'(\rho,t)$ solution of this equation is assumed to be accurately described by a small perturbation around the fixed point solution $u'(\rho,t)=u_*'(\rho)+\delta u'(\rho,t)$, hence
\beq
\begin{aligned}\label{linp}
\partial_t \delta u' =& - 2(u_*'+\delta u')+(d-2)\rho(u_*'+\delta u')'\\
&- (u_*'+\delta u')' F(u_*'+\delta u'),
\end{aligned}
\eeq
where $F(u')=1/(1+u')^2$ is introduced, and $\partial_t u_*'$ vanishes by definition. Expanding it around the fixed point solution up to linear order gives 
\beq
F(u_*'+\delta u') \approx F(u_*')+\frac{\partial F}{\partial u'}(u'_*)\delta u'.
\eeq
Thus, considering the last term in \eqref{linp}
\beq
\begin{aligned}
(u_*'+\delta u')' F\approx&(u_*'+\delta u')'\left[F(u_*')+\frac{\partial F}{\partial u'}(u'_*)\delta u'\right]\\
=&u_*''F(u_*')+\delta u'' F(u_*')+u_*''\frac{\partial F}{\partial u'}(u'_*)\delta u',
\end{aligned}
\eeq
where the last term coming from the product in the right-hand side is neglected since the perturbation assumed to be small. The solution of the fixed point equation satisfies
\bea\label{fpeq}
- 2u_*' + (d-2)\rho u_*''-u_*''F(u_*')=0,
\eea
hence
\bea
	\partial_t \delta u' = - 2\delta u'+(d-2)\rho\delta u''
	- \delta u'' F\left(u_*'\right)-u_*''\frac{\partial F}{\partial u'}\left(u'_*\right) \delta u',\nn
\eea
which can be recast into the following form
\beq
\begin{aligned}
\partial_t \delta u' =& - 2\delta u'+(d-2)\rho\delta u''
- \delta u''\left(\rho(d-2)-2\frac{u_*'}{u_*''}\right) \\
&-\delta u'\left((d-2)-2\frac{\partial}{\partial \rho}\frac{u_*'}{u_*''}\right),
\end{aligned}
\eeq
where the relation $\partial F/\partial u'=1/u'' \partial F/\partial \rho$ is used and $F(u_*')$ is expressed from \eqref{fpeq}. Further manipulating the right-hand side gives
\beq
\partial_t \delta u'= - d \delta u'+2 \frac{u_*''^2 -u_*'u_*'''}{u_*''^2}\delta u'+2\frac{u_*'}{u_*''}\delta u'',
\eeq
which after some algebra provides the final result for the fluctuation equation
\beq\label{pde}
\partial_t\delta u' = 2 \frac{u_*'}{u_*''}\left(\partial_\rho - \frac{(u_*' u_*'')'}{u_*'u_*''}-\frac{d-4}{2}\frac{u_*''}{u_*'} \right)\delta u'.
\eeq
The solution of the PDE in \eqref{pde} is found by using the method of separation of variables, that is $\delta u'=f(t)g(\rho)$. A straightforward computation gives
\beq
f(t)\propto \exp{\theta t} \qquad\text{and} \qquad g(\rho)\propto u_*'^{\frac{1}{2}(\theta+d-2)}u_*''.
\eeq
Thus, the complete solution up to a constant factor
\beq
\delta u'(t,\rho)\propto e^{\theta t} u_*'^{\frac{1}{2}(\theta+d-2)}u_*'',
\eeq
where $\theta\in\mathbb{R}$ is given by the regularity condition described in the text.  
\bibliographystyle{apsrev}

\end{document}